\documentclass[%
 reprint,showpacs,
 amsmath,amssymb,
 aps,nofootinbib,superscriptaddress
]{revtex4-1}
 \usepackage{ulem}
    \usepackage{float}
\usepackage{graphicx}
\usepackage{dcolumn}
\usepackage[colorlinks=true, linkcolor = red, citecolor = blue]{hyperref}
\usepackage{subcaption}
\usepackage{float}
\usepackage{xcolor}

\begin{document}
\newcommand{\lu}[1]{{#1}}
\newcommand{\quita}[1]{\textcolor{orange}{#1}}
\newcommand{\JCH}[1]{\textcolor{blue}{#1}}
\title{Production of PBHs from inflaton structures}

\author{Juan Carlos Hidalgo}
  \email{hidalgo@icf.unam.mx}
\affiliation{Instituto de Ciencias Físicas, Universidad Nacional Autónoma de México,
62210, Cuernavaca, Morelos, México.}

\author{Luis E. Padilla}
\email{lepadilla@icf.unam.mx}
\affiliation{Instituto de Ciencias Físicas, Universidad Nacional Autónoma de México,
62210, Cuernavaca, Morelos, México.}
\affiliation{Mesoamerican Centre for Theoretical Physics,
Universidad Aut\'onoma de Chiapas, Carretera Zapata Km. 4, Real
del Bosque (Ter\'an), Tuxtla Guti\'errez 29040, Chiapas, M\'exico.}

\author{Gabriel German}
  \email{gabriel@icf.unam.mx}
  \affiliation{Instituto de Ciencias Físicas, Universidad Nacional Autónoma de México,
62210, Cuernavaca, Morelos, México.}

\date{\today}%

\begin{abstract}
At times prior to big bang nucleosynthesis, the Universe could show a primordial structure formation period if dominated by a fast oscillating inflaton field during reheating. In this context, we have postulated a new mechanism of primordial black hole formation  [L. E. Padilla, J. C. Hidalgo, and K. A. Malik, Phys. Rev. D, \textbf{106}, p. 023519, Jul 2022], that draws on the analogy between an extended reheating era and the scalar field dark matter model, contemplating the gravitational collapse of inflaton halos and inflaton stars. In this paper we look at the requirements for the realization of this new mechanism. We show that a generic primordial power spectrum with a peak at small scales is most suitable for the production of a considerable number of primordial black holes (PBHs). When such a requirement is met, and if reheating lasts long enough, large populations of PBHs with $M_{\rm PBH}\sim 1~\mathrm{gram}$ may be produced. We find, in particular, that the mass fraction of PBHs is orders of magnitude larger than that obtained when PBHs form via direct collapse in a Universe dominated by radiation or pressureless dust. Looking at observable implications of our findings, we explore the possibility that \lu{Planck mass relic remnants of the evaporation of the PBHs could contribute to be the totality of the dark matter in the universe.}
 \end{abstract}

\maketitle

\section{Introduction}

In recent years, primordial black holes (PBHs) have gained considerable attention due to the recent detection by LIGO and Virgo of black holes in the intermediate mass range \citep{abbott2020gw190521}. A possible explanation for these events could be the merger product of two PBHs \citep{sasaki2016primordial}. However, PBHs have been widely studied for the last 50 years \citep{zel1967hypothesis}, given that these objects could be responsible for a great variety of physical phenomena. For example, PBHs with masses $M_{\rm PBH}\leq 10^{15}~\rm{g}$ are expected to have evaporated at the present time due to Hawking radiation \citep{hawking1974black}, with the possibility of leaving behind a supersymmetric particle or a Planck mass relic (see for example \citep{carr2010new}), which in turn could be a dark matter component. At the other  side of the spectrum, in which  $M_{\rm PBH} > 10^{15}~\rm{g}$,  these objects and their populations are relevant in several cosmological and astrophysical phenomena (for a detailed description of all these effects, see e.g.~\citep{carr2010new,carr2020constraints}). 

If black holes are indeed formed in the early Universe, then a natural question is whether they (or their remnants) could constitute part or the totality of the dark matter component of the Universe. PBHs could form in a wide range of masses, and appear in models of the early universe at a single mass or in extended spectra. Currently, only a few values of PBH masses could still contribute to the totality of dark matter. In particular, if remnants are left over after PBH evaporation,  the tiny particles could still constitute all of the dark matter.

The standard scenario for the formation of PBHs suggests that after inflation, primordial density perturbations stretched beyond causal contact during the inflationary era, reentering the horizon in subsequent stages as matter inhomogeneities. If the amplitude of these fluctuations is large enough, then they can   collapse under their own gravity into PBHs \citep{zel1967hypothesis,hawking1971gravitationally}. The criteria for this to occur have been determined by detailed numerical simulations \citep{Escriva:2019phb,Musco:2004ak,Bloomfield:2015ila,Escriva:2021aeh}, and are mostly expressed in terms of a threshold value $\delta_c$ for the density contrast, evaluated at the time of cosmological horizon crossing. In the standard big bang scenario, in which the Universe is radiation-dominated after inflation, this threshold value is roughly $\delta_c^{(\rm rad)}\simeq 0.41$ (see for example \citep{harada2013threshold}).

In most inflationary models, the transition from the inflationary period to the radiation dominance epoch would not be expected to occur immediately. In fact, it is expected that, after inflation, the inflaton rolls quickly to the minimum of its potential and oscillates around this minimum while it transfers energy to other (standard model) fields. This process is generically dubbed reheating. The reheating process may last for a few $e$-foldings of expansion and the energy scale of this period is bound only by the big bang nucleosynthesis (BBN) process (at about $10~\rm{MeV}$).  

In a significant part of the reheating process, the inflaton field oscillates at the minimum of a potential that can be generically approximated by a quadratic potential. In this period the Universe thus presents a dustlike dominance, where fluctuations yield a primordial structure formation process. In recent work \citep{Niemeyer:2019gab,PhysRevD.103.063525,Eggemeier:2021smj}, it has been  proposed that this primordial structure formation process is analogous to the structure formation period in the so-called scalar field dark matter (SFDM) model \cite{2014NatPh..10..496S,PhysRevD.95.043541,Urena-Lopez:2019kud,10.1007/978-3-319-02063-1_9}, and thus, phenomena associated to  the SFDM model may occur during this extended reheating scenario. For example, galaxylike haloes may form, referred as {inflaton halos} or inflaton clusters, as well as the associated solitonic corelike structures at the center of each halo, referred to as {inflaton stars} in reheating. 

{The formation of supermassive black holes has been proposed in the SFDM model as a result of the gravitational collapse of the central soliton/core in galaxies \citep{Padilla:2020sjy}.} 
Following the analogy drawn above, two of us proposed in a previous paper a new mechanism for the formation of PBHs during the reheating epoch, through the gravitational collapse of massive enough inflaton haloes and inflaton stars \citep{Padilla:2021zgm}. We found that in each case there is a critical threshold value of the primordial density contrast under which structures should gravitationally collapse to form PBHs. Such thresholds can be an order of magnitude below that associated to collapse during the radiation era\footnote{In the same work, we  explored the possibility of including the effects of a quartic attractive self-interaction. In that case, the collapse threshold value $\delta_c$ can be further reduced by a few orders of magnitude.}. {Later, in \citep{DeLuca:2021pls} this scenario was extended by considering the effects of accretion on the primordial structures, contributing to reach the critical mass at which PBHs are formed.}

As it is well known, in order to produce PBHs during either the radiation or the reheating periods, the primordial power spectrum (PPS) requires an enhancement at small scales to amplitudes of order $\sim 10^{-4}-10^{-2}$. 
This is because the amplitude favored by the cosmic microwave background (CMB) at the pivot scale $k_\star=0.05~\rm{Mpc^{-1}}$ is constrained to be $\mathcal{P}_\mathcal{R}(k_\star) =2.10\times 10^{-9}$ \citep{Planck:2018jri}, which is too low for a significant number of PBHs to be produced. 
There are a few methods proposed in the literature that could achieve this small-scale enhancement.
 For example, peaks in the PPS could be reached in multifield hybrid models of inflation (see for example \citep{Garcia-Bellido:1996mdl}). In this class of models, one of the fields acts as the inflaton while a second field becomes responsible for generating the peak in the PPS towards the end of inflation. Another proposal is to undergo a phase of ultraslow-roll inflation, in which the inflaton reaches an inflection point in the potential \citep{Tsamis:2003px,Kinney:2005vj,Iacconi:2021ltm}. The duration of the ultraslow-roll phase and the location of the inflection point define in turn the location and amplitude  of the peak in the PPS. Another realization of the PPS enhancement has been explored in the Higgs inflationary scenario, where a noncanonical kinetic term and a noncanonical coupling of the Higgs field result in a peak of the PPS \citep{Lin:2021vwc}. {In relation with the above, it has been shown that in single field models of inflation the PPS can grow as fast as $k^{5}(\log k)^2$ \cite{Carrilho:2019oqg} (see also \citep{Byrnes:2018txb}).} In all these scenarios, if the peak of the PPS occurs very close to the end of inflation, it pertains scales that reenter the horizon during reheating. 

\begin{figure}
    \includegraphics[width=3.7
    in]{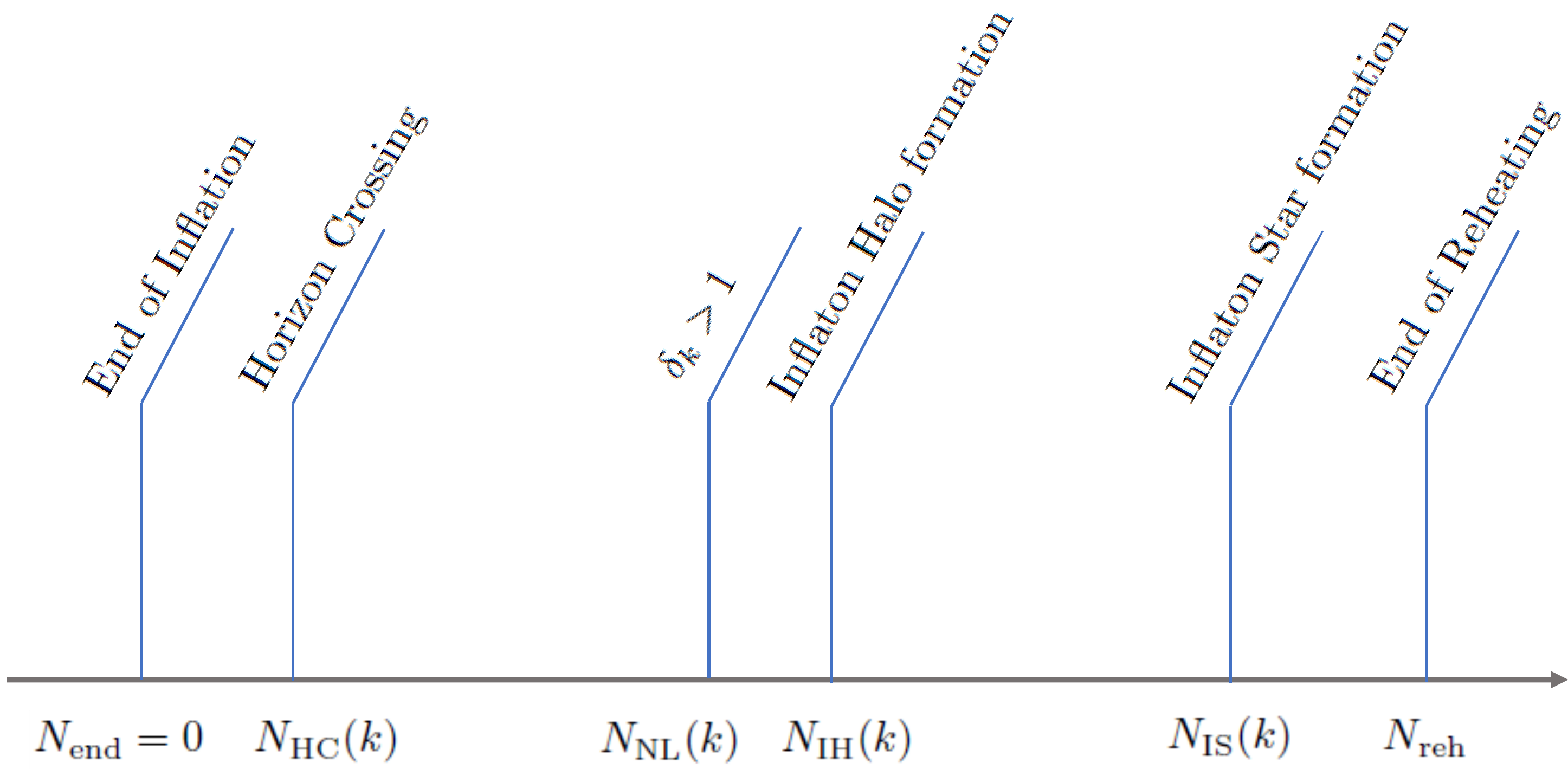}
    \caption{\footnotesize{Stages of evolution of density fluctuations during reheating in terms of $e$-foldings after the end of inflation ($N_{\rm end}$). A mode of wavenumber $k$ enters the horizon $N_{\rm HC}(k)$ $e$-folds after reheating starts, then reaches a non-linear amplitude at $N_{\rm NL}(k)$, forming an inflaton halo a Hubble time later, at $N_{\rm IH}(k)$. After the virialization of such halo, a condensation of the inflaton at the core produces a central soliton, an inflaton star, at  $N_{\rm IS}(k)$. Thermalization is expected to end reheating at $N_{\rm reh}$. If this last event occur earlier, the sequence is interrupted and only some or none of the $k$-dependent processes may take place.}}
    \label{fig:description}
\end{figure}

The main motivation of the present paper is to work out the details of PBH production in light of the mechanism presented in \citep{Padilla:2021zgm}, compute PBH population abundances, and study the fate of the produced black holes. To preserve the generality of our study, we consider a generic PPS with a peak at the smallest scales without looking at a particular inflationary model. The sequence of events we seek to characterize, from the end of inflation up to PBH formation, is illustrated in Fig.~\ref{fig:description}. A detailed account of each stage is described below.


Our paper is organized as follows.
{In Sec. \ref{sec:II} we present a parametrization of the PPS, suitable to produce virialized structures during reheating. The process of structure formation and the resulting structures are described in Sec.~\ref{subsec:IIIA}. The possibility that PBHs form from these structures is evaluated in Sec.~\ref{subsec:IIIB}. This is presented in contrast with the direct collapse of primordial fluctuations during reheating, accounted for in Sec.~\ref{subsec:IVA}, as well as the case when the adopted potential is realized in an instant reheating scenario, and PBHs form during the radiation era (Sec.~\ref{subsec:IVB}). The abundance of PBHs produced in the considered scenarios is presented in Sec.~\ref{sec:V} where we discuss the production efficiency among different mechanisms. With the mass spectrum at hand, the fate of the evolved black hole populations is analyzed in Sec.~\ref{sec:VI}, in terms of the possible dominance of the energy budget by PBHs and Planck mass relics as leftovers of the evaporation process. Our concluding remarks are presented in Sec.~\ref{sec:VII}}.

\section{A suitable Power spectrum at small scales}
\label{sec:II}

\subsection{The primordial curvature power spectrum}
\label{subsec:IIA}

During the inflationary era, it is typically assumed that a single scalar field $\varphi$ slowly rolls down its potential $V(\varphi)$, generating an epoch of accelerated expansion. During this time, the quantum fluctuations of the inflaton are stretched out of the Hubble horizon and converted to classical perturbations. This is also manifest in metric perturbations at the level of curvature perturbations and primordial gravitational waves. In the simplest scenario, within the slow-roll regime, the PPS of curvature perturbation can be approximated as
\begin{equation}\label{pps_eps}
    \mathcal{P}_\mathcal{R}(k) = \frac{H^2}{8\pi^2M_{\rm Pl}^2\epsilon},
\end{equation}
where $H$ is the Hubble parameter, $M_{\rm Pl}$ is the Planck mass, and $\epsilon$ is the first slow-roll parameter.\footnote{The first two slow-roll parameters $\epsilon$ and $\eta$ are defined in terms of derivatives of the potential as $\epsilon\equiv (1/2) [V'(\varphi)/V(\varphi)]^2$ and $\eta \equiv |V''(\varphi)/V(\varphi)|$, with a prime denoting derivative with respect to $\varphi$. We assume that inflation lasts up to the moment when $\epsilon = 1$.} The above expression can be compared to the following, which is parametrized in terms of observables of the CMB, 
\begin{equation}\label{pps1}
    \mathcal{P}_\mathcal{R}(k) = \mathcal{A}_s\left(\frac{k}{k_\star}\right)^{n_s-1},
\end{equation}
where $\mathcal{A}_s$ is the amplitude of the perturbations, typically quoted at the pivot scale $k_\star = 0.05~\rm{Mpc^{-1}}$, and $n_s$ is known as the spectral index which, in a first approximation, is considered scale independent. The Planck 2018 TT+LowE+lensing data constrains this form of the potential at the 95\% C.L. and at the pivot scale $k_*$ as \citep{Planck:2018jri}
\begin{equation}
    \ln{(10^{10}A_s)} = 3.044\pm 0.014,\ \ \ \ n_s = 0.9634\pm 0.0048.\label{const}
\end{equation}

\begin{figure}[t!]
    \centering
    \includegraphics[width=3.3in]{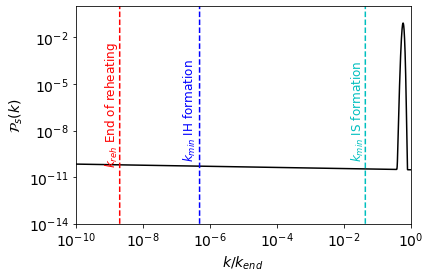}
    \caption{\footnotesize{PPS \eqref{pps_gauss} for the fiducial values $k_p = 0.6\cdot k_{\rm end}$, $\Sigma_p = 0.03\cdot k_{\rm end}$, and $\mathcal{B}_s=0.084$. The red, blue, and cian lines indicate the minimum $k$-modes that are relevant for the reheating epoch, the formation of inflaton halos, and the formation of inflaton stars, respectively (a more in depth discussion is presented below in section \ref{sec:III}).}}
    \label{fig:pps}
\end{figure}

In view of these values, a substantial production of PBHs during the early evolution of the Universe requires an enhancement of the PPS at the smallest scales, away from those relevant to the CMB. {Since our motivation is to characterize our new mechanism of PBH formation proposed in \citep{Padilla:2021zgm}, in this work} we propose a simple parametrization of the PPS with a generic peak at small scales
\begin{equation}\label{pps_gauss}
       \mathcal{P}_\mathcal{R}(k) = \mathcal{A}_s\left(\frac{k}{k_\star}\right)^{n_s-1}+\mathcal{B}_s\exp\left[-\frac{(k-k_p)^2}{2\Sigma_p^2}\right],
\end{equation}
{i.e.}~we include a Gaussian peak located at $k_p$ and with a variance $\Sigma_p^2$. The amplitude of the peak in the PPS is controlled by the auxiliary amplitude parameter $\mathcal{B}_s$. Throughout this paper, we shall work with this PPS, adopting the mean values of Eq.~\eqref{const}. Specific realizations of inflationary potentials which yield such PPS will be left for later work. Here, as a generic example, we work with the particular values $k_p = 0.6 \cdot k_{\rm end}$, $\Sigma_p = 0.03\cdot k_{\rm end}$, and $\mathcal{B}_s=0.084$\footnote{\lu{These particular values were chosen so that figure \ref{fig:beta} could show clearly the abundance of PBHs in each of the scenarios studied. For smaller values of the peak, the abundance of PBHs in some of the other scenarios would be completely negligible.}}. The PPS generated for these parameters is shown in Fig.~\ref{fig:pps}.

\subsection{The amplitude of primordial density fluctuations}
\label{subsec:IIB}
For definiteness, let us fix relevant parameters of the generic inflationary model, with the purpose of producing PBHs significantly from the PPS proposed in the last section. 


{A model-independent bound for the Hubble parameter at horizon crossing follows from considering the recent bound ($r<0.032$) for the tensor-to-scalar ratio \citep{Tristram:2021tvh}. From Eqs.~\eqref{pps_eps} and \eqref{pps1} we have, at $k=k_*$,
\begin{equation}
    H_* = \sqrt{\frac{A_sr}{2}\pi M_{\rm Pl}}\leq 4.44\times 10^{13}~\rm{GeV},
\end{equation}
where $A_s\simeq 2.1\times 10^{-9}$. An estimate of the value of $k_{\rm end}$ follows by assuming that at the end of inflation $H_{\rm end}\approx H_{*}$ and that the spectral index does not change much from its value at $k=k_*$ thus,
\begin{equation}
A_s(k_{\rm end})\left(\frac{k_{\rm end}}{k_*}\right)^{n_s-1}= \frac{H_{*}^2}{8\pi^2M_{\rm Pl}^2},
\end{equation}
where we have set $\epsilon=1$  and we have dropped the second term in Eq.~\eqref{pps_gauss}, since it does not contribute much at $k_{\rm end}$. The term on the left can be approximated following
\begin{eqnarray}
\left(\frac{k_{\rm end}}{k_*}\right)^{n_s-1}&=&\left(\frac{a_{\rm end}H_{\rm end}}{a_*H_*}\right)^{n_s-1}\nonumber \\
&\approx& \left(\frac{a_{\rm end}}{a_*}\right)^{n_s-1}=e^{N_* (n_s-1)},
\end{eqnarray}
where $N_*$ is the number of $e$-folds during inflation. For $50<N_*<60$ the value is $e^{N_*(n_s-1)} \approx 0.14$. It therefore follows that $A_s(k_{e})\approx 3\times 10^{-11}$ and that $k_e\approx 0.346\ \rm{m^{-1}}.$}
This in turn fixes, via the Friedmann equation $H^2 = \rho/(3 M_{\rm Pl}^2)$, the energy scale at which inflation ends:
{
\begin{equation}
    \rho_{\rm end} = (1.368\times 10^{16}\ \rm{GeV})^4.
\end{equation} }

Once we have a prescription for the PPS at hand, we can look at the matter density fluctuations, in order to test for the critical amplitudes for the collapse. In the comoving gauge, we can express $\mathcal{P}_\delta$ in terms of the PPS $\mathcal{P}_\mathcal{R}$ as \citep{Wands:2000dp}
\begin{equation}
    \mathcal{P}_\delta(k,t) = \left[\frac{2(1+\omega)}{5+3\omega}\right]^2\left(\frac{k}{aH}\right)^4\mathcal{P}_\mathcal{R}(k,t),
\end{equation}
where $\omega$ is the equation of state of the dominating component. This is evaluated at the moment of horizon reentry, where the mean amplitude of matter density fluctuations is given by
\begin{equation}
    \bar\delta_{\rm HC}(k) \equiv \left(\frac{\delta\rho}{\rho}\right)_{ k= aH} = \sqrt{\mathcal{P}_\mathcal{\delta}(k)}.
\end{equation}

Subsequently, the amplitude will grow to form structures during reheating as we describe in the following.



\section{Structure formation during reheating}
\label{sec:III}

\subsection{Primordial structure formation period}
\label{subsec:IIIA}
After the inflationary epoch, it is expected the inflaton rolls quickly to the minimum of its potential and presents fast oscillations until decay, when it transfer its energy to the rest of the particles of the standard model of particle physics (see e.g.~\cite{Lozanov:2019jxc} for a comprehensive review). During this period, most potentials can be approximated as
\begin{equation}\label{V_eff}
    V(\varphi) = \frac{1}{2}\mu^2\varphi^2+...
\end{equation}
where $\mu^2\equiv d^2V(\varphi)/d\varphi^2|_{\varphi_{\rm min}}$ 
 and $\varphi_{\rm min}$ is the value of the field at the minimum of the potential, which is usually taken equal to zero, i.e. $V(\varphi_{\rm min})=0$. 

Although the value of $\mu$ depends on the particular potential $V(\varphi)$, a fast oscillating regime demands the condition $\mu\gg H$. We meet such condition by taking the specific value $\mu = 10 H_{\rm end}$ and assuming an immediate transition from the end of inflation to the fast oscillations (a reasonable approximation for this potential is shown in \cite{Carrion:2021yeh}).

In our approximation the quadratic potential controls the oscillations, and consequently the cosmological background is expected to go through a dustlike evolution after the inflationary epoch:
\begin{equation}\label{background}
    \rho(a) \simeq \rho_{\rm end}\left(\frac{a_{\rm end}}{a}\right)^3.
\end{equation}
This stage of the Universe can last for a considerable number of $e$-folds, up to the inflaton decay, which could take place as late as BBN, at roughly  $\rho_{\rm BBN}\sim (1\times 10^{-2}\ \rm{GeV})^4$.

As we already mentioned, perturbations at the small scales of the spectrum can reenter the horizon during this phase. For a given scale $k$, the number of $e$-folds at horizon reentry $N_{\rm HC}(k)$, after the end of inflation, is given by
\begin{equation}\label{eq:hc}
    N_{\rm HC}(k) = 2\ln\left(\frac{k_{\rm end}}{k}\right).
\end{equation}
After horizon reentry, perturbations grow as $\delta\sim a$ and may become nonlinear. The number of $e$-folds required to reach a nonlinear regime is a function of the wave number,
\begin{equation}\label{Eq:NNL}
    N_{\rm NL}(k) = N_{\rm HC}(k)+\ln[1.39\delta_{\rm HC}^{-1}(k)]. 
\end{equation}
{In the above expression $\delta_{\rm HC}$ is the amplitude of matter density fluctuations at horizon crossing.} 

Once inhomogeneities reach a nonlinear amplitude, inflaton halos are expected to form within a Hubble time \cite{Niemeyer:2019gab,Padilla:2021zgm}. In terms of $e$-folds this occurs at
\begin{equation}\label{N_ih}
    N_{\rm IH}(k) =N_{\rm NL}(k)+ \frac{2}{3}\ln\left(1+\frac{H^{-1}}{t_{\rm NL}(k)}\right),
\end{equation}
where $t_{\rm NL}(k) = [2/(3H_{\rm end})]\cdot [e^{N_{\rm HC}(k)}1.39/\delta_{\rm HC}(k)]^{3/2}$. 

Ultimately, if reheating lasts long enough to reach $t_{\rm IS}(k) \simeq t_{\rm NL}(k)+\Delta t_{\rm cond}(k)$, an inflaton star forms at the core of the inflaton halo. Here $\Delta t_{\rm cond}(k)$ accounts for the time required for an inflaton star to condensate at the center of the formed halo. The condensation process starts once the inhomogeneity becomes non-linear and is given by
\begin{equation}\label{t_cond}
    \frac{\Delta t_{\rm{cond}}(k)}{t_{\rm NL}(k)} = 8.168\times 10^{-18}\left({\mu_5^2}{M_{\rm Pl}^2}M_{\rm IH}(k)R_{\rm IH}(k)\right)^{3/2},
\end{equation}
where in the above expression $\mu_5\equiv \mu/(10^{-5}M_{\rm Pl})$. The condensation time then sets the $e$-folds required for inflaton stars to form:
\begin{equation}\label{N_is}
    N_{\rm IS}(k) =N_{\rm NL}(k)+\frac{2}{3}\ln\left(1+\frac{\Delta t_{\rm cond}(k)}{t_{\rm NL}(k)}\right).
\end{equation}

In Eq. \eqref{t_cond} $M_{\rm IH}(k)= (4\pi M_{\rm Pl}^2/H_{\rm end})\cdot(k_{\rm end}/k)^3$ is the mass of the inflaton halo, which we equate with the mass of the cosmological horizon at the horizon crossing time, that is,
\begin{equation}
    \left(\frac{M_{\rm IH}(k)}{7.1\times 10^{-2}~\rm{g}}\right) =\left(\frac{1.8\times 10^{15}~\rm{GeV}}{H_{\rm end}}\right)\left(\frac{k_{\rm end}}{k}\right)^3,
\end{equation}
and $R_{\rm IH}(k) = [3M_{\rm IH}(k)/(4\pi\cdot 200\rho(a_{\rm NL}))]^{1/3}$ is its radius. Additionally, the mass with which inflaton stars are expected to form meets the condition
\begin{equation}\label{mcmh}
    \left(\frac{M_{\rm IS}(k)}{2.4\times 10^{-5}~\rm{g}}\right) = \frac{\rho_{11}^{1/6}(a_{\rm NL})}{\mu_5}\left(\frac{M_{\rm IH}(k)}{7.1\times 10^{-2}\ \rm{g}}\right)^{1/3},
\end{equation}
with $\rho_{11}(a)\equiv 200\rho(a)/(10^{11}\ \rm{GeV})^4$. Equivalently, we can rewrite the above expression using Eqs. \eqref{background}, \eqref{Eq:NNL}, and the relation $\rho(a_{\rm NL}) = \rho(a_{\rm HC})\cdot (a_{\rm HC}/a_{\rm NL})^3$ as
\begin{equation}
    \left(\frac{M_{\rm IS}(k)}{2.4\times 10^{-5}~\rm{g}}\right) = \sqrt{\frac{\delta_{\rm HC}(k)}{1.39}}\frac{\rho_{11}^{1/6}(a_{\rm HC})}{\mu_5}\left(\frac{M_{\rm IH}(k)}{7.1\times 10^{-2}\ \rm{g}}\right)^{1/3}.
\end{equation}
From the previous expression we can see then that the mass of inflaton stars strongly depends on the value of $\delta_{\rm HC}(k)$.

{The primordial structure formation period can be thus summarized as follows: In an extended reheating era, and a few $e$-foldings after horizon reentry, the primordial matter perturbations evolve and become nonlinear [as dictated by Eq.~\eqref{Eq:NNL}]. Then virialization is ensued, and a halo is formed after only a Hubble time. Subsequently, and only if reheating lasts long enough for the condensation to occur at the core of the halos, a soliton dubbed inflaton star is formed at the center of each halo (the sequence is illustrated in Fig.~\ref{fig:description}). We emphasize that, if reheating is interrupted early enough, only some or none of the processes in the primordial structure formation may take place.}

\subsection{PBHs from the collapse of inflaton structures}
\label{subsec:IIIB}

If at the horizon crossing time the perturbations were dense enough, then the primordial structures could collapse to form PBHs. In a recent work \citep{Padilla:2021zgm}, we determined the threshold values in case the inflaton halo or inflaton star collapsed into PBHs. These are, respectively,
\begin{equation}
    \delta_c^{(\rm IH)} = 0.238, \ \ \ \ \ \delta_c^{(\rm IS)} = 0.019.
\end{equation}
 In the featured mechanism the time for PBHs formation should be of the same order as the formation times presented above for the inflaton halos and inflaton stars. It is clear that, for a perturbation with a wave number $k$ to collapse to form a PBH during reheating, its collapsing time should take place before the end of reheating, here parametrized by $N_{\rm reh}$. Mathematically, this condition is 
\begin{equation}
    N_{\rm reh}\geq N_{i}(k), \ \ \ \ i = {\rm IH,\, IS}.
\end{equation}

\noindent Thus, PBH formation during reheating is modulated by the parameters $N_{\rm reh}$ and $N_{i}(k)$, which in turn are closely related to $k$ and $\delta_{\rm HC}(k)$. 

Note that, during the collapse to form PBHs, it is expected that dissipation and accretion processes take place, and therefore the final mass of the PBH may not be equivalent to the mass of the progenitor structure. One can parametrize this difference by expressing the formation mass as $M^{i}_{\rm PBH}(k) = \gamma_{i}M_{i}(k)$ ($i=\rm{IH},\ \rm{IS}$), where $\gamma_i$ denotes the efficiency of the collapse. The exact value of $\gamma_i$ should be fixed through detailed numerical calculations, which are still under way for the case of reheating (partial progress has been reported in \cite{deJong:2021bbo,Padilla:2021uof}). Intuitively, since during this phase of the evolution
the inflaton is expected to behave effectively as dust, we might expect no great resistance to the collapse (except for a quantum pressure effect on the de Broglie wavelength scales associated to the inflaton), which leads us to expect that $\gamma_i\simeq O(1)$. Thus we adopt the approximation $\gamma_i= 1$. 

To illustrate the relevance of the set of scales mentioned above, in Fig.~\ref{fig:pps} we sketch with vertical lines the largest structures (smallest $k$) that reenters the cosmological horizon during reheating (in red), the largest scale that can form inflaton-halo/PBH structures (in blue), and the largest scale that could form inflaton stars/PBHs (in cyan) for our  model and the particular value $N_{\rm reh} = 40$. \lu{The smallest $k$ scales are simply obtained by replacing in the left hand side of Eqs. \eqref{eq:hc}, \eqref{N_ih}, and \eqref{N_is} the value $N_{\rm reh} = 40$.} As we can see, inflaton halos show a  much broader spectrum than inflaton stars, as would the PBHs formed from the collapse of these two types of structures. The reason for this is the time the inflaton takes to condense in the center of inflaton halos, which is much larger than the time required to form the inflaton haloes themselves. Therefore, in order to form PBHs from the collapse of the inflaton stars, reheating must last long enough and  the enhancement of the PPS (the peak) must appear very close to the end of inflation. 

\begin{figure}
    \centering
    \includegraphics[width=3.5in]{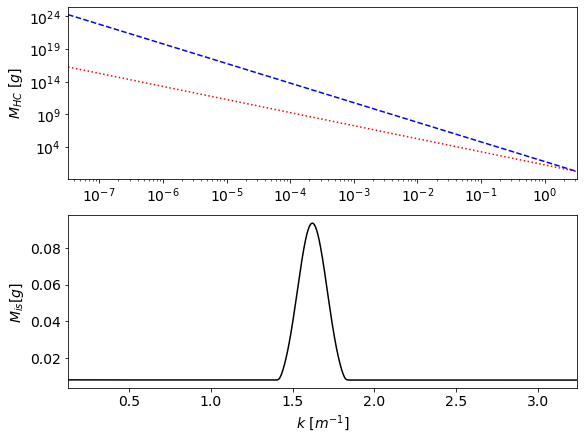}
    \caption{\footnotesize{Top panel: The mass of the cosmological horizon at the horizon crossing time as a function of $k$ for an universe dominated by the inflaton during reheating (blue dashed line) and a universe dominated by a radiation component (red dotted line). Bottom panel: The mass spectrum of inflaton stars, expressed in  Eq.~\eqref{mcmh}, as a function of $k$.}}
    \label{fig:mass_spect}
\end{figure}

As a complement, in Fig.~\ref{fig:mass_spect} we show the mass spectrum of the inflaton structures, as a function of $k$ modes.\footnote{{To obtain Figs.~\ref{fig:pps} and \ref{fig:mass_spect} we used the simplification $\delta_{\rm HC}(k) = \bar\delta_{\rm HC}(k)$. We adopt this same simplification when calculating quantities related to inflaton stars. In a more accurate approximation to the mass spectrum, $\delta_{\rm HC}(k)$ should follow a Gaussian distribution, with  configurations collapsing from a critical, and larger, amplitudes, according to the Press-Schechter formalism. Thus  in this case, our approximation represents an overestimation of the average mass of Inflaton Stars (since in this work $\sqrt{\mathcal{B}_s}>\delta_c^{\rm (IS)}$). This however suffices for the proof of principle we intend to demonstrate in the present work.}} As justified above, for inflaton haloes the mass is that of the cosmological horizon at the horizon crossing time, while the mass of the PBHs formed from inflaton stars is much smaller and breaks the one-to-one correspondence with $k$ modes. This brings interesting consequences for the resulting PBH abundance, as we describe below. The distinctive mass spectrum for inflaton stars can be attributed to its dependence on the nonlinear scale; namely,  $M_{\rm IS}(k)\sim \rho^{1/6}(a_{\rm NL})M_{\rm IH}^{1/3}(k)$ [see Eq. \eqref{mcmh}]. {For comparison, we also display the mass spectrum of PBHs formed in the standard collapse during reheating or a radiation-dominated era.}

\section{Comparison with existent scenarios of PBH formation}
\label{sec:IV}

\subsection{PBH formation via direct collapse during reheating}
\label{subsec:IVA}

The formation of PBHs during reheating has been previously modeled as pressureless dust configurations collapsing without limit (see e.g.~Refs.~\citep{Khlopov:1980mg,1982AZh....59...15P} for pioneering works). In this scenario, primordial perturbations that reenter the cosmological horizon are expected to collapse gravitationally and it is the deviation from spherical symmetry what modulates the probability of PBH formation. {This is because, if perturbations are made of nonrelativistic matter, deviations from sphericity lead to a pancake collapse}. Through the Hoop conjecture, Harada \textit{et al.}~\citep{Harada:2016mhb} showed that only the most spherically symmetric perturbations are able to form PBHs while the rest of the perturbations { end up as virialized objects} (forming inflaton halos). Subsequently, Ref.~\citep{Harada:2017fjm} showed that the effect of angular momentum becomes important when the amplitude of the perturbations at the horizon crossing time is small enough, obtaining a significant suppression in the production of PBHs. 

Note that the size and evolution stages of perturbations that later form inflaton haloes are common to the early stages of PBH formation via direct collapse.
The PBH formation criteria in the collapse of haloes and in the direct collapse are not mutually exclusive and instead both should be taken as complementary. If at the moment of direct collapse the initial perturbation was sufficiently nonspherical, it will be destined to virialize in a process of violent relaxation. In any case, if the initial perturbation is massive enough, then the formation of a PBH is inevitable.

\subsection{Standard PBH formation scenario: Instantaneous reheating}
\label{subsec:IVB}
For comparison, we evaluate the PBH production for the derived PPS of Eq.~\eqref{pps_gauss} in the standard radiation-dominated universe, taking reheating as instantaneous. In this scenario, perturbations that reenter the cosmological horizon directly collapse to form PBHs, but a larger threshold amplitude $\delta_c^{(\rm rad)}$ is required. 
In a radiation background, perturbations with wave number $k$ are expected to reenter the cosmological horizon at $    N_{\rm HC}^{(\rm rad)}(k)$ $e$-folds after the end of inflation. This is given by
\begin{equation}
    N_{\rm HC}^{(\rm rad)}(k) = \ln\left(\frac{k_{\rm end}}{k}\right).
\end{equation}

Afterwards, and if the amplitude of the contrast density evaluated at the horizon crossing time is large enough, overdensities collapse into PBHs with a mass close to that of the cosmological horizon at the time of horizon crossing:
\begin{equation}
    \left(\frac{M_{\rm PBH}^{(\rm rad)}(k)}{7.1\times 10^{-2}~\rm{g}}\right) =  \gamma_{\rm rad}\left(\frac{1.8\times 10^{15}~\rm{GeV}}{H_{\rm end}}\right)\left(\frac{k_{\rm end}}{k}\right)^2,
\end{equation}
where $\gamma_{\rm rad} = (1/3)^{3/2}$ is the collapse efficiency parameter. 

The precise value of the threshold of the density contrast at the horizon crossing time $\delta_c^{(\rm rad)}$ has been the subject of many studies (see for example, \citep{Niemeyer:1997mt,shibata1999black,Musco:2008hv,musco2013primordial,Harada:2013epa}), and it is subject to the characteristics of curvature profile in the initial perturbation \citep{Musco:2018rwt,Nakama:2013ica,Escriva:2021pmf} and, in particular, on how compact the perturbation was at the time of reentering the cosmological horizon. Here we adopt the most accepted   value of $\delta_c^{(\rm rad)} = 0.41$.

In Fig.~\ref{fig:mass_spect} we plotted the mass of the cosmological horizon at the horizon crossing time for each of the scales in our example model. As we can see, the mass contained at the horizon crossing time in a radiation-dominated universe is smaller than in the case of the reheating scenario. Such characteristic may help distinguishing the nature of PBHs in an eventual detection.
In the following we evaluate the abundance of PBHs in the three scenarios discussed.  

\section{Abundance of primordial black holes}
\label{sec:V}

To compute the mass fraction of PBHs we employ the usual Press-Schechter formalism \citep{Press:1973iz}, where the probability that a particle at \textbf{x} is part of a collapsed object with mass $>M$ is equivalent to the probability that a density field smoothed on some scale $R$ -- which corresponds to a mass scale $M$ -- exceeds the threshold value $\delta_c$:
\begin{equation}\label{Pdelta}
    P[\delta>\delta_c] = \int_{\delta_c}^{\infty}P(\tilde\delta)d\tilde\delta.
\end{equation}
Here $P(\delta)$ represents the probability density function associated to $\delta$. We assume a Gaussian distribution
\begin{equation}
    P(\delta) = \frac{1}{\sqrt{2\pi}\sigma(R)}\exp\left(-\frac{\delta^2}{2\sigma(R)^2}\right),
\end{equation}
where $\sigma(R)$ is the standard deviation of $\delta$ \lu{evaluated at the horizon crossing time}, 
\begin{equation}
    \sigma(R
    )^2 = \int_0^\infty W^2(\tilde kR)\mathcal{P}_\delta(\tilde k,t_{\rm HC})d\ln \tilde k,
\end{equation}
$W(kR) = \exp(-k^2R^2/2)$ is the Fourier transform of the window function used to smooth the density contrast over a scale $R = 1/k$, and $\mathcal{P}_\delta$ is the power spectrum of density perturbations. It is worth mentioning that the maximum $\sigma$ reached from our proposed power spectrum is approximately   $\sigma_{\rm max} =  2\times 10^{-2}$.

Plugging the above expressions in Eq.~\eqref{Pdelta} we obtain
\begin{equation}\label{Pdelta2}
    P[\delta>\delta_c] = \frac{1}{2}\text{erfc}\left(\frac{\delta_c}{\sqrt{2}\sigma(R)}\right).
\end{equation}

The mass fraction $\beta(M)$ is defined such that $\beta(M)d\ln M$ corresponds to the fraction of the Universe comprised of structures with masses between $M$ and $M+dM$. Consequently, $\int_M^\infty\beta(M)d\ln M$ corresponds to the fraction of objects in the Universe with masses larger than $M$. Noticing that in the case the mass $M$ of the objects can be expressed as $M = 4\pi\rho R^3/3$, this allows us to compute the fraction of the total energy density collapsing into objects of mass $M$ as
\begin{equation}\label{betaM}
    \beta(M) = -2M\frac{\partial R}{\partial M}\frac{\partial P[\delta>\delta_c]}{\partial R},
\end{equation}
where the factor 2 is included to fit estimations from the peaks theory. 

\begin{figure}
    \centering
    \includegraphics[width=3.5in]{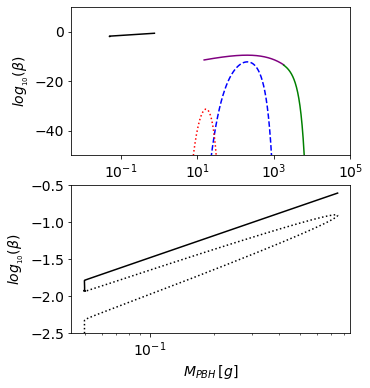}
    \caption{\footnotesize{Top panel: The density fraction $\beta$ as a function of $M_{\rm PBH}$ for the collapse of inflaton stars (black), inflaton haloes (dashed blue) and radiation overdensities (dotted red). The direct collapse of overdensities in reheating is determined by the sphericity criterion (purple) and the contribution of angular momentum (green). Bottom panel: A zoom of the abundance of PBHs collapsed from inflaton stars, with two contributions to each mass (dotted black line) produced from different $k$-modes (see Fig.~\ref{fig:mass_spect}). The solid line is the addition of both contributions.}}
    \label{fig:beta}
\end{figure}

In principle, to compute the fraction of the total energy density collapsing into PBHs of a given mass $M_{\rm PBH}$ due to the collapse of inflaton structures or radiation overdensities, we should just compute the expression in Eq. \eqref{betaM}. The results obtained for our model are plotted in Fig.~\ref{fig:beta}. However, the collapse of inflaton stars requires a special treatment. \lu{We can compute the abundance of PBH due to the collapse of inflaton stars simply by calculating the expression
\begin{equation}\label{eq:beta_is}
    \beta (M_{\rm IS}) =     \frac{M_{\rm IS}}{M_{\rm IH}}\beta^{\rm (IS)}(M_{\rm IH}),
\end{equation}
where in the above formula
\begin{equation}
    \beta^{\rm (IS)}(M_{\rm IH})\equiv-2M_{\rm IH}\frac{\partial R}{\partial M_{\rm IH}}\frac{\partial P[\delta>\delta_c^{\rm  (IS)}]}{\partial R},\nonumber
\end{equation}
and the reader must consider that the previous expression is only calculated for those structures that do manage to form an inflaton star in their center, i.e. for larger $k$'s than the one marker by the cian line in Fig. \ref{fig:pps}. Notice that once we calculate the abundance of PBHs due to the collapse of inflaton stars as showed in Eq. \eqref{eq:beta_is}} we \lu{obtain} the formation of PBHs with the same mass for two different $k$ modes, as illustrated in Fig.~\ref{fig:mass_spect}. Thus the Press-Schechter formula \eqref{eq:beta_is} should not be applied directly. The simplest way to account for the two contributions to a single mass, is by computing each contribution for each mass of the PBHs formed separately, and then add the contributions to each mass together, keeping in mind that these PBHs are formed at different times. 

The case of the direct collapse in a dustlike reheating scenario is also computed through formulas beyond Eq.~\eqref{betaM}. Such case is limited by the sphericity requirement described in Sec.~\ref{subsec:IVA}. The density fraction for that case is related to the mean amplitude of perturbations as \citep{Harada:2016mhb} 
\begin{equation}
\label{beta:sphere}
     \beta \simeq 0.05556\sigma^5,\quad \text{for} \quad 0.005\lesssim \sigma\lesssim 0.2.
\end{equation}

\noindent In the regime $\sigma< 0.005$ the effects of angular momentum of primordial fluctuations become important. The consequence is that PBH production is modified as \citep{Harada:2017fjm}
\begin{equation}
\label{beta:I}
    \beta \simeq 1.9\times 10^{-7}f_q(q_c)\mathcal{I}^6\sigma^2 \exp(-0.15\mathcal{I}^{4/3}\sigma^{-2/3}),
\end{equation}

\noindent where $\mathcal{I}$ is a parameter of order $O(1)$ and $f_q(q_c)$ is the fraction of mass with a level of quadrupolar asphericity $q$ smaller than a threshold $q_c$ (following Ref.~\citep{Harada:2017fjm}, we take $f_q(q_c),\ \mathcal{I} = 1$ in our estimations). 

Our results are plotted in Fig.~\ref{fig:beta}, where we present the $\beta(M)$ values from the possibilities of PBH formation described above. Most prominent of the top panel is the black solid line, which is the density fraction of inflaton stars collapsed onto PBHs. The lower panel shows the two contributions to $\log_{\rm 10}\beta(M)$ when considering the collapsed inflaton stars. The density fraction is unusually large in this case, but that intentionally expected from the amplitude of the variance chosen in our example (see the figure of Ref.~\citep{Padilla:2021zgm}).

Next in the top panel of Fig.~\ref{fig:beta} are the purple and green lines which represent the density fraction of the direct collapse in a dustlike era, expressed in Eqs.~\eqref{beta:sphere} and \eqref{beta:I}.  The blue dashed line is the abundance of PBHs resulting from the direct collapse of inflaton haloes, while the red dotted line shows the production of collapsed overdensities in a radiation era if reheating is instantaneous.  

Figure~\ref{fig:beta} is the main result of the present work, so it is worth discussing these results in more depth. Note first that the mass range for PBHs differs for each mechanism, with the smallest masses resulting from the inflaton star collapse, and the largest from the direct collapse of dustlike overdensities during reheating, a mechanism that encompasses the mass range of PBHs formed if reheating was instantaneous, and if PBHs were formed from inflaton halo collapse. 

Moreover, as expected, forming PBHs in a reheating dust-like period is much easier than in a radiation period. For our PPS with a localized peak, the production of PBHs via direct collapse during reheating is several orders of magnitude higher than during radiation. Specifically,  we observe that the production of PBHs due to the direct collapse during reheating, even when bound by sphericity and to low values of spin, is a more efficient mechanism, with $\beta_{\rm max}^{\rm (dir)} = 1.4\times 10^{-10}$, than that due to the gravitational collapse of the inflaton halos ($\beta_{\rm max}^{\rm (IH)} = 5.8\times 10^{-33}$). This is however, dependent on the value of $\sigma$; at  values $\sigma \gtrsim 5\times 10^{-2}$ the abundance of collapsed inflaton haloes is expected to dominate over the direct collapse result (recall that for our working example, $\sigma_{\rm max} = 2\times 10^{-2}$).

As is evident also from the figure, the production of PBHs due to the collapse of inflation stars is a more efficient mechanism than any of the alternatives, with $\beta_{\rm max}^{\rm (IS)} = 0.14$. And the least efficient production mechanism is through collapse during radiation if reheating is instantaneous ($\beta^{\rm (rad)}_{\rm max} = 1.6\times 10^{-80}$). All this is true for our example values of the variance with $\sigma \gtrsim 2.4\times 10^{-3}$. Below this value the production from direct collapse could be more efficient. In hindsight it is evident that our choice of the peak amplitude is taken so we can show in a single plot the resulting abundances of all the mechanisms reviewed in this paper. 


\section{Subsequent evolution: Evaporation of PBHs}
\label{sec:VI}

After formation, small PBHs start losing mass via the Hawking radiation mechanism. The rate at which the mass of PBHs decreases due to this effect is given by \citep{1974Natur.24830H}
\begin{equation}\label{teva}
  M_{\rm PBH}(t) = M_{\rm PBH}(t_{f})\left(1-\frac{t-t_f}{\Delta t_{\rm eva}}\right)^{1/3},   
\end{equation}
where $t_f$ is the time of formation and $\Delta t_{\rm eva}$ is the time at which a PBH evaporates completely. This is given by,
\begin{equation}
    \Delta t_{\rm eva} \equiv t_{\rm eva}-t_f=t_{\rm Pl}\left(\frac{M_{\rm PBH}(t_{f})}{M_{\rm Pl}}\right)^{3},
\end{equation}
  where $t_{\rm Pl}$ is the Planck time. Two possibilities have been proposed as a result of PBH evaporation. Either PBHs evaporate completely or a remnant particle may survive. In the first case, PBHs with masses $M_{\rm PBH}\lesssim 10^{15}\ \rm{g}$ vanish by the present time, and only more massive PBHs could be found in the Universe today. For the mass spectrum presented in Fig.~\ref{fig:beta} the mass of  PBHs is so small, and we would expect them to evaporate way before primordial nucleosynthesis (BBN). This opens the possibility that PBHs could contribute to reheat the early Universe \citep{Domenech:2020ssp,Domenech:2021wkk}. Such possibilities and the associated constraints to specific inflationary models will be explored elsewhere (this possibility has been considered previously in \citep{Hidalgo:2011fj,PhysRevD.54.6040,Zagorac2019GUTscalePB,Martin:2019nuw}). The alternative, a second possibility,  which emerges in the context of quantum gravity (see for example \citep{COLEMAN1992175}), suggests that black hole evaporation stops when the mass of the black hole reaches the Planck mass, leaving behind a relic that may contribute to the dark matter of the Universe. In what follows we will elaborate more on this second scenario.

\begin{figure}
    \centering
    \includegraphics[width=3in]{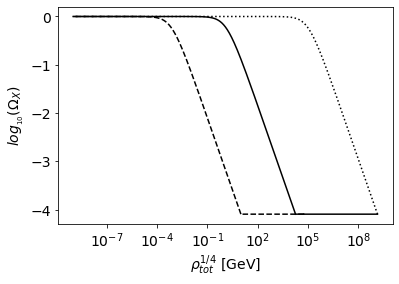}
    \caption{\footnotesize{Mass fraction of PBHs and Planck mass relics as a function of $\rho_{\rm tot}^{1/4}$ for an extended reheating scenario. Dotted, solid and dashed lines corresponds to $N_{\rm reh} = 25$, $N_{\rm reh} = 40$, and $N_{\rm reh} = 55$, respectively.}}
    \label{fig:omegas}
\end{figure}

\begin{figure}
    \centering
    \includegraphics[width=3in]{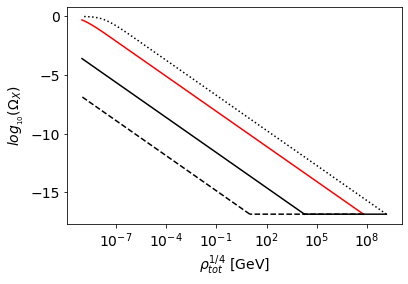}
    \caption{\footnotesize{Same as figure \ref{fig:omegas} but with the parameter $\mathcal{B}_s = 0.084/50$. The red solid line corresponds to $N_{\rm reh} = 29.1925$.}}
    \label{fig:omegas_2}
\end{figure}

To be concrete, we look at the contribution of nonevaporated PBHs plus Planck mass relics to the dark matter, at a given time (for example, at some time previous to BBN). Taking $\bar\beta$ as the mass fraction in absence of Hawking radiation, the evolved mass fraction of PBHs plus Planck mass relics at time $t$ is given by  \citep{Martin:2019nuw}:
\begin{eqnarray}\label{omx}
    \Omega_{X}(t) =&& \int_{\hat M_{\rm min}}^{\hat M_{\rm max}}\bar\beta(M_{\rm PBH},t)\left(1-\frac{t-t_f}{\Delta t_{\rm eva}}\right)^{1/3}d\ln M_{\rm PBH}\nonumber \\
    + && \int_{\tilde m_{\rm min}}^{\tilde m_{\rm max}}\bar\beta(M_{\rm PBH},t)\frac{m_{\rm Pl}}{M_{\rm PBH}}d\ln M_{\rm PBH},
\end{eqnarray}
where $\hat M_{\rm min}$ ($\hat M_{\rm max}$) is the minimum (maximum) mass of PBHs formed that have not evaporated at time $t$, whereas $\tilde m_{\rm min}$ ($\tilde m_{\rm max}$) is the minimum (maximum) mass with which PBHs that evaporated by that time originally formed. As shown in \citep{Martin:2019nuw}, we can express $\bar\beta(M_{\rm PBH},t) = b(t)\bar\beta(M_{\rm PBH},t_{\rm ref})$, where $t_{\rm ref}$ is a reference time, and $b(t)$ fulfills the differential equation 
\begin{equation}
    \dot b(t)+\left(\frac{\dot \rho_{\rm tot}}{\rho_{\rm tot}}+3H\right)b(t) = 0.
\end{equation}
Here $\rho_{\rm tot}$ is the total energy density of the background universe. During reheating we expect to have only the inflaton field and PBHs, which means that we only have components that behave like nonrelativistic matter. In that case, during this period we would have that $b(t) = \mathrm{const}$. This result suggests that we can take the formation time as the reference time, that is $t_{\rm ref} = t_{\rm f}$, $b(t) = 1$, and $\bar\beta(M_{\rm PBH},t) = \bar\beta(M_{\rm PBH},t_{\rm f})$ for times prior to thermalization. Once reheating ends, $b(t)$ should start evolving according to
\begin{equation}
    \frac{db}{d\ln \rho_{\rm tot}}+\frac{\Omega_X-1}{\Omega_X-4}b = 0,
\end{equation}
where $\rho_{\rm tot} =\rho_{\rm rad}+\rho_{X}$, and with $\rho_{X}$ the energy density of PBHs plus Planck mass relics. Finally, we can compute $\Omega_{X}$ from Eq.~\eqref{omx} as a function of cosmic time, which may evolve in a nontrivial way. During reheating, we expect the cosmic time to evolve according to a universe filled completely with nonrelativistic matter, i.e. $t\sim a^{3/2}$, whereas after reheating 
\begin{equation}
    \frac{d(t-t_f)}{d\ln \rho_{\rm tot}} = \frac{\sqrt{3}m_{\rm Pl}}{(\Omega_X-4)\sqrt{\rho_{\rm tot}}}.
\end{equation}
We plotted in Fig.~\ref{fig:omegas} the evolution of $\Omega_X$ as a function of $\rho_{\rm tot}$ for our generic model and for three cases; from left to right, $N_{\rm reh} = 25,\ 40$, and $55$, respectively. \lu{As we can see, for our example  we obtain that in all cases the universe should be dominated by Planck mass relics prior to matter-radiation equality (at around $(10^{-9}~\rm{GeV})^4$) which raises the question of whether the model meet the cosmological constraints.}  
\lu{Even in the limiting case in which $N_{\rm reh} = 55$ (being very close to the maximum number of $e$-folds that reheating can last if it happened in a dust-like background \citep{German:2022sjd}) we find that the universe would inevitably be dominated by Planck mass relics due to the gravitational collapse of the inflaton stars}. This is because during a radiation-dominated universe $b(t) \propto a(t)$ and because the mechanism of formation of PBHs from inflaton stars turns out to be very efficient.   \lu{In fact the model and realizations presented here can be tested through cosmological constraints. Let us, for example, limit the amplitude of the PPS peak to  $\mathcal{B}_s = 0.084/50$. We plot in Fig. \ref{fig:omegas_2} the evolution of $\Omega_X$ as a function of $\rho_{\rm tot}$ for this new set of parameters. The line drawn in red corresponds to the limit value of reheating where the PBHs could constitute the totality of the dark matter in the universe.}
In this sense, our mechanism is capable of  constraining the size of the peak in the PPS (so that PBHs are not overproduced through the collapse of inflaton stars) given a duration of the reheating period, or vice versa.

\section{Conclusions and discussion}
\label{sec:VII}
In this article we have explored the realization of a recently proposed mechanism for the formation of PBHs, via the collapse of inflaton structures during reheating. We compare the efficiency of PBH production from our mechanism with a standard formation scenario of direct collapse both during reheating and in the radiation-dominated era (the case of instantaneous reheating).
The possibility that this mechanism is efficient depends strongly on the duration of the reheating period and the position and amplitude of the peak in the PPS. We found that a localized peak (with maximum variance $\sigma_{\rm max}^2 =  4\times 10^{-4}$), close to the end of inflation, plus a large number of reheating  $e$-folds, are required to produce PBHs significantly from the collapse of inflaton stars.

Our main result is summarized in Fig.~\ref{fig:beta}. We have shown that the formation of PBHs in our mechanism is not only delayed with respect to the standard scenario, but also modifies the mass at which a PBH is formed, for a prescribed $k$ mode. In particular, the elements that define the mass of an inflaton star result in two different formation periods (and two different associated wave numbers) for PBHs of a given mass. This, together with the low threshold that overdensities required to form PBHs, produces a significant abundance of these objects at the low-mass end of the spectrum. We have shown that, if reheating lasts long enough, PBH production via the collapse of inflaton stars is about nine orders of magnitude more efficient than the direct collapse of overdensities during reheating, which in turn is more efficient (for low values of the variance) than the collapse of inflaton haloes.

Since most of the PBHs studied here are bound to evaporate through Hawking radiation, we have followed the evolution of populations of PBHs through this process. In particular, the possibility that they could get to dominate the energy density of the Universe if Planck mass relics are leftover after evaporation. This and the alternative that PBHs could reheat the Universe by the mechanism presented here will be explored in more depth once specific inflationary models are worked out. \lu{Considering a new set of parameters for our PPS we also constraint the duration that the reheating epoch should last in order for the Planck mass relics generated after the evaporation of the tiny PBHs contribute to be the totality of the dark matter in the Universe.}

The large number of primordial black holes that can be formed from our mechanism compared to the case of instantaneous reheating motivates us to search in future for models of inflation that could be constrained through this new scenario of PBH formation.
The particular characteristics of the mechanism explored here can help to identify extended reheating through the usual observables of PBHs. This is a pending task that will be explored in subsequent works.




\begin{acknowledgments}
{The authors  acknowledge support from program UNAMPAPIIT, Grants No. IN107521 “Sector Oscuro y Agujeros Negros Primordiales” and No. IG102123 ``Laboratorio de Modelos y Datos (LAMOD) para proyectos de Investigación Científica: Censos Astrofísicos". L.E.P. and J.C.H. acknowledge sponsorship from CONACyT
Network Project No.~304001 ``Estudio
de campos escalares con aplicaciones en cosmolog\'ia y
astrof\'isica'', and through Grant No. CB-2016-282569.  The work of
L.E.P. is also supported by the DGAPA-UNAM postdoctoral grants program, by CONACyT M\'exico under Grants No. A1-S-8742, No. 376127 and
FORDECYT-PRONACES Grant No. 490769.}
\end{acknowledgments}

\bibliographystyle{ieeetr}
\bibliography{biblio}
\end{document}